\documentclass[aps,prb,twocolumn,groupedaddress,amssymb,amsmath,showpacs]{revtex4}
\usepackage{dcolumn}
\usepackage{graphicx}

\begin{document}

\title{Topological defects in graphene: dislocations and grain boundaries}

\author{Oleg V. Yazyev}
\affiliation{Department of Physics, University of California, Berkeley, California 94720, USA}
\affiliation{Materials Sciences Division, Lawrence Berkeley National Laboratory, Berkeley, California 94720, USA}
\author{Steven G. Louie}
\affiliation{Department of Physics, University of California, Berkeley, California 94720, USA}
\affiliation{Materials Sciences Division, Lawrence Berkeley National Laboratory, Berkeley, California 94720, USA}

\date{\today}


\begin{abstract}
Topological defects in graphene, dislocations and grain boundaries, 
are still not well understood despites the considerable number of 
experimental observations. We introduce a general 
approach for constructing dislocations in graphene characterized by 
arbitrary Burgers vectors as well as grain boundaries, covering the 
whole range of possible misorientation angles. By using {\it ab initio}
calculations we investigate thermodynamic and electronic properties 
of these topological defects, finding energetically favorable symmetric 
large-angle grain boundaries, strong tendency towards out-of-plane 
deformation in the small-angle regimes, and pronounced effects on the 
electronic structure. 
The present results show that dislocations and grain boundaries
are important intrinsic defects in graphene which may be used for 
engineering graphene-based nanomaterials and functional devices.
\end{abstract}

\maketitle

\section{INTRODUCTION}

The isolation of graphene, a two-dimensional (2D) material with 
extraordinary physical properties, has opened new horizons for 
physics exploration and future technology.\cite{Geim07,Katsnelson07} 
In 2D, properties of materials can be heavily affected by structural 
irregularities. Graphene edges and point defects such as vacancies have 
been extensively investigated over the past few years.\cite{CastroNeto09} 
However, these types of disorder have to be distinguished from
dislocations and grain boundaries, structural defects characterized
by the finite values of their respective topological invariants,
Burgers vectors and misorientation angles.\cite{Nelson02} Such topological 
defects introduce non-local disorder into the crystalline lattice. 
Surprisingly, dislocations and grain boundaries in graphene are still not 
well understood despite the growing number of experimental observations. 
 
The first experimental results date back to the scanning 
tunneling microscopy (STM) studies of tilt grain boundaries on graphite 
surfaces,\cite{Albrecht88} fueled by their confusion with biological 
macromolecules.\cite{Clemmer91,Heckl92} More recently it has been 
shown that grain boundary defects have a dramatic influence on the 
local electronic properties of graphite.\cite{Cervenka09,Cervenka09b} 
An individual dislocation in free-standing graphene layers has been 
imaged using transmission electron microscopy (TEM).\cite{Hashimoto04}
Topological defects resulting from either kinetic factors
or substrate imperfections have also been reported for epitaxial
graphene grown on SiC,\cite{Miller09} Ir(111)\cite{Coraux08,Loginova09} and 
polycrystalline Ni surfaces.\cite{Park10}

Here, we describe a systematic approach for constructing 
arbitrary dislocations and grain boundaries in graphene starting
from disclinations as the elementary topological defects. Then, by
using {\it ab initio} calculations we explore energetic and electronic
properties of the proposed structures finding a number of intriguing 
features such as two energetically favorable symmetric 
large-angle grain boundaries, strong tendency towards out-of-plane 
deformation in the small-angle regimes, and pronounced effects on the 
electronic structure. Our results highlight the possible important role of 
dislocations and grain boundaries in practical graphene samples.

The present paper is organized in the following manner. In Section~\ref{methods} 
we describe our first-principles computational methodology.
Section~\ref{constructions} presents a systematic approach for constructing atomic 
structures of dislocations and grain boundaries in graphene. 
Sections~\ref{energetics} and \ref{electronic} are devoted to the discussion
of energetics and electronic structure of the constructed topological defects,
respectively. Section~\ref{conclusions} concludes our work.

\section{COMPUTATIONAL METHODS}\label{methods}

\begin{figure}
\includegraphics[width=8.5cm]{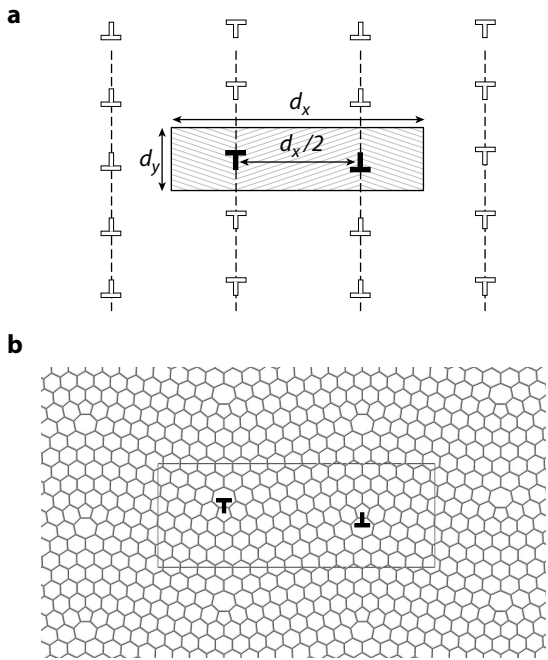}
\caption{\label{figs3}
(a) Schematic illustration of the $d_x \times d_y$ rectangular simulation 
supercell with two dislocations (filled symbols) separated by $d_x/2$.
The periodic images of the dislocations are shown as empty symbols.
The dashed lines depict the grain boundary lines. (b) One of
grain boundary models ($\theta = 9.4^\circ$) used in the present study. 
The simulation supercell is indicated.
}
\end{figure}

First-principles calculations have been performed
using the spin-polarized density functional theory (DFT) scheme
implemented in the \texttt{SIESTA} code.\cite{SIESTA} The generalized
gradient approximation (GGA) exchange-correlation density functional
\cite{Perdew96} was employed together with a double-$\zeta$ plus
polarization basis set, norm-conserving pseudopotentials
\cite{Troullier91} and a mesh cutoff of 200~Ry. The computational 
model involved two parallel equally spaced grain boundaries 
in a rectangular simulation supercell in order to satisfy periodic 
boundary conditions (see Fig.~\ref{figs3}). The distance between the 
neighboring dislocations along the boundary line, and thus the 
misorientation angle $\theta$, are changed by varying the $d_y$ 
supercell dimension. The $d_x$ supercell dimension was $\sim$4~nm in
all studied models. The chosen supercell construction allows one to reduce
the error due to elastic interactions between the neighboring grain 
boundaries.\cite{Blase00} We verified that a larger inter-boundary 
separation ($d_x = 8$~nm) produces only a negligible change in the calculated 
grain boundary energy in both small-angle and large-angle regimes. 
Both atomic coordinates and supercell dimensions were optimized using 
the conjugate-gradient algorithm and a 0.04\ eV/\AA\ maximum force 
convergence criterion. The Brillouin zone was sampled using 2 $k$-points
along the $x$ axis and a consistent number of approximately
$8/d_y$ $k$-points along the $y$ axis ($d_y$ in nm). 
The scanning tunneling microscopy (STM) images were simulated using a previously developed 
method\cite{Meunier99,Ishigami04} employing the calculated local 
density-of-states in the energy window of $\pm 0.6$~eV around the charge neutrality point.

\section{DISCUSSION OF RESULTS}\label{discussion}

\subsection{Atomic structure of topological defects: \\
a systematic approach}\label{constructions}

\begin{figure}
\includegraphics[width=8.5cm]{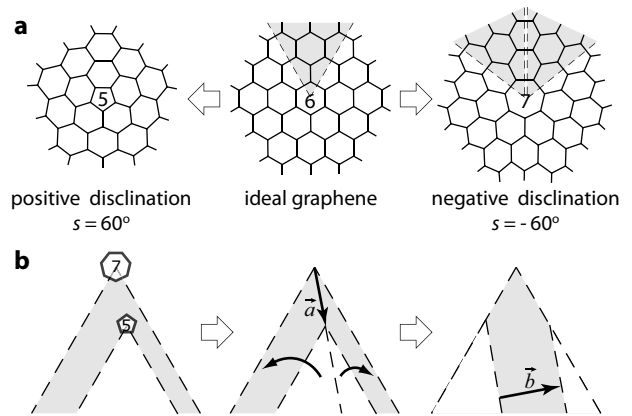}
\caption{\label{fig1}
(a) Positive ($s=60^\circ$) and negative ($s=-60^\circ$) 
disclinations in graphene are produced by either removing or adding 
a $60^\circ$ wedge (shaded area) of material without 
changing the coordination of carbon atoms. (b) A pair of 
complementary disclinations is equivalent to a dislocation: a 
negative disclination inserts a $60^\circ$ wedge while a positive 
disclination removes such a wedge within one of the seven equivalent 
sectors. The introduced amount of material (shaded area) can also 
be viewed as a semi-infinite strip of width $|\vec{b}|$. 
}
\end{figure}

In truly 2D materials only edge dislocations are possible since the Burgers 
vector $\vec{b}$, a topological invariant which reflects the magnitude and 
direction of the crystalline lattice distortion produced by a dislocation, 
is constrained to lie in the material's plane. One can imagine such 
dislocation as a result of embedding a semi-infinite strip of width 
$|\vec{b}|$ into an otherwise perfect 2D crystalline lattice. As a 
guiding rule for constructing atomic structures of dislocations in 
graphene we assume that the dislocation core is free from under- or 
over-coordinated carbon atoms; that is, we aim at minimizing the 
energy of the dislocation core and, thus, the total formation energy
of the dislocation.\cite{Hirth92} To develop such construction we 
adopt a membrane theory approach which views a dislocation as a pair 
of positive and negative disclinations, i.e. topological defects obtained
by removing and adding a semi-infinite wedge of material to an 
otherwise perfect crystalline lattice, respectively.\cite{Seung88} As shown
in Figure~\ref{fig1}(a), $s=60^\circ$ ($s=-60^\circ$) disclination in 
graphene contains a five (seven) membered ring in its core while the 
original three-fold coordination of all carbon atoms is preserved. 
Figure~\ref{fig1}(b) schematically shows the equivalence of a pair of 
complementary disclinations to a dislocation. Moreover, we 
find that on graphene lattice the distance between two disclinations, 
$|\vec{a}|$, is related to the resulting Burgers vector $\vec{b}$ by a simple 
relation, $|\vec{a}|=|\vec{b}|$ (for proof see Appendix~\ref{AppA}). 

\begin{figure*}
\includegraphics[width=17cm]{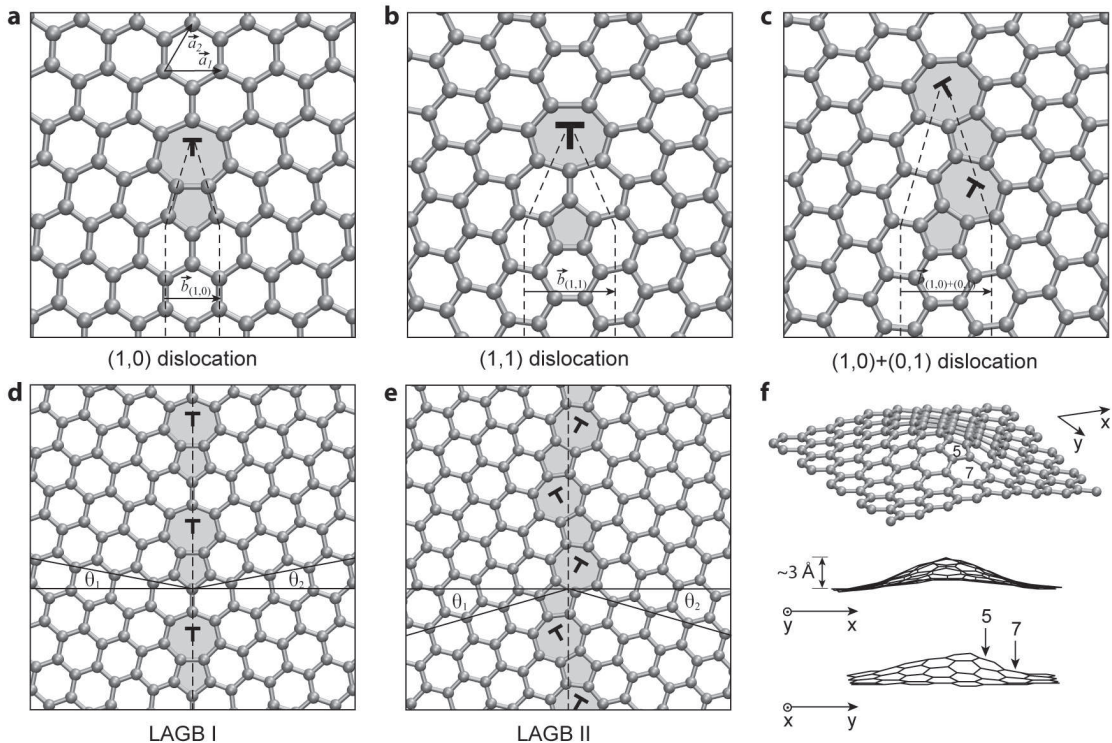}
\caption{\label{fig2} 
(a-c) Atomic structures of (1,0) and (1,1) dislocations, and
a (1,0)+(0,1) dislocation pair, respectively. 
The dashed lines delimit the introduced semi-infinite strips of graphene
originating at the dislocation core. Non-6-membered rings are shaded.
(d,e) Atomic structures of the $\theta = 21.8^\circ$ (LAGB~I)
and the $\theta = 32.2^\circ$ (LAGB~II) symmetric large-angle grain boundaries,
respectively. The dashed lines show the boundary lines and the solid lines
definite angles $\theta_1$ and $\theta_2$.  
(f) Buckling of the graphene layer due the presence of a (1,0) dislocation. 
}
\end{figure*}

Since any Burgers vector $\vec{b}$ is a proper translational vector of 
graphene lattice, i.e. $\vec{b}=n\vec{a}_1+m\vec{a}_2$ 
($\vec{a}_{1,2}=(3d_{\rm cc}/2, \pm \sqrt{3} d_{\rm cc}/2)$;
$d_{\rm cc} = 1.42$\ \AA, the nearest neighbor interatomic distance in
graphene), we will use the pair of integers $(n,m)$ as a descriptor of
dislocations in graphene. This notation is analogous to the chirality
indices used to describe the structure of carbon nanotubes. The core 
of the shortest Burgers vector dislocation (1,0) 
($|\vec{b}_{\rm (1,0)}|=\sqrt{3}d_{\rm cc}=2.46$~\AA) contains an 
edge-sharing heptagon-pentagon pair as shown in Figure~\ref{fig2}(a). The
$(1,0)$ dislocation inserts a semi-infinite strip of atoms along the 
armchair high-symmetry direction in graphene while its Burgers
vector is oriented along the zigzag direction. This simplest dislocation
structure has been extensively studied in the context of plastic deformation 
of carbon nanotubes,\cite{Nardelli98} nanotubes junctions\cite{Chico96} as 
well as graphene itself.\cite{Ewels02,Carpio08} 
The second member of the family, the (1,1) dislocation, has a larger 
Burgers vector ($|\vec{b}|=3d_{\rm cc}=4.23$~\AA) and inserts a 
semi-infinite strip along the zigzag direction of graphene (see Fig.~\ref{fig2}(b)). 
Alternatively, the core of the dislocation with the same Burgers vector
can be constructed from two $|\vec{b}_{\rm (1,0)}|=2.46$~\AA\ dislocations, 
(1,0) and (0,1), e.g. as shown in Figure~\ref{fig2}(c). The simple method 
outlined above can be used to build dislocation with even longer Burgers 
vectors, inevitably leading to larger elastic energies.

Grain boundaries, the interfaces between the domains of material with 
different crystallographic orientations, are commonly viewed as periodic 
arrays of dislocations.\cite{Read50} Particularly, in 2D materials such as graphene, 
one-dimensional (1D) chains of edge dislocations constitute tilt 
grain boundaries. Mutual orientation of the two crystalline domains is 
described by the misorientation angle $\theta = \theta_1 + \theta_2$ ($\theta \in (0^\circ,60^\circ)$ 
in graphene), a topological invariant defined as shown in Fig.~\ref{fig2}(d). 
Another parameter $\psi = |\theta_1 - \theta_2| \in (0^\circ,\theta)$ describes the 
inclination of the boundary line with respect to the symmetric 
configuration ($\psi = 0^\circ$). We limit our consideration to only 
symmetric ones since asymmetric configurations tend to result in diverging 
elastic energies.\cite{Carraro93} Importantly, due to the presence of 
two high-symmetry directions in graphene, armchair and zigzag, both 
misorientation angles close to 0$^\circ$ and 60$^\circ$ can be 
considered as small-angle grain boundaries along these two directions, 
respectively. Aligning (1,0) dislocations along the grain boundary 
line results in a discrete set of misorientation angles $\theta$ in 
accordance with Frank's equation\cite{Hirth92}
\begin{equation}
 \theta = 2 \arcsin \frac{|\vec{b}_{(1,0)}|}{2d_{(1,0)}},
\end{equation}
where $d_{(1,0)}$ is one of the possible values for the distance 
between the neighboring dislocations. Large values of $d_{(1,0)}$ 
correspond to small-angle grain boundaries along the armchair direction. 
The closest possible packing of (1,0) dislocations results in the 
large-angle grain boundary structure shown in Fig.~\ref{fig2}(d). This 
configuration characterized by $\theta = 21.8^\circ$ (LAGB~I)
has already been suggested in the literature.\cite{Heckl92,Simonis02}
In order to cover the the range of $\theta$ between 21.8$^\circ$ and 
60$^\circ$, it is necessary to introduce another type of dislocations,
e.g. (1,1) dislocations:
\begin{equation}
 \theta = 60^\circ - 2 \arcsin \frac{|\vec{b}_{(1,1)}|}{2d_{(1,1)}}.
\end{equation}
The smallest value of $d_{(1,1)}$ gives rise to the LAGB~I structure 
rotated by 180$^\circ$. Large separations $d_{(1,1)}$ correspond to 
small-angle grain boundaries along the zigzag direction. Alternatively,
small-angle grain boundaries along this direction can be constructed 
using the (1,0)+(0,1) pairs with the densest possible packing of 
dislocations leading to the structure LAGB~II with $\theta = 32.2^\circ$ 
(Fig.~\ref{fig2}(e)).
Hence, it is possible to construct symmetric grain boundaries covering
the whole range of $\theta$ from 0$^\circ$ to 60$^\circ$ by using (1,0) 
dislocations and either (1,1) dislocations or (1,0)+(0,1) dislocation 
pairs. We stress that the present construction is equally applicable to
tilt grain boundaries in both graphene and graphite.

\subsection{Energetics of topological defects}\label{energetics}

\begin{figure}
\includegraphics[width=8.5cm]{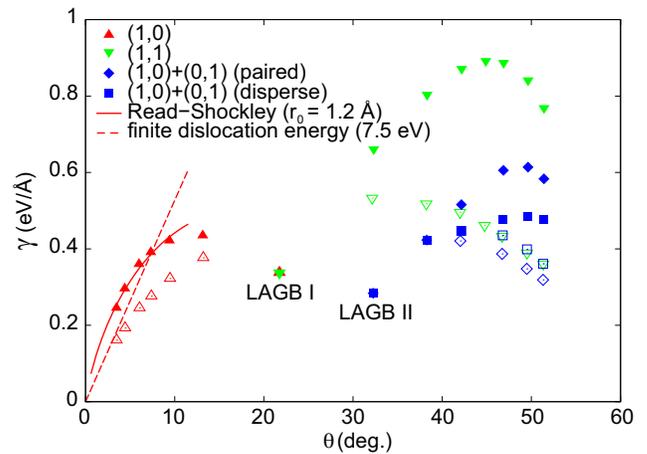}
\caption{\label{fig3} 
(color online).
Grain boundary energy per unit length $\gamma$ as a function 
of misorientation angle $\theta$ for various flat (filled symbols) and
buckled (open symbols) grain boundary structures. The two energetically
favorable large-angle grain boundaries, LAGB~I and LAGB~II, are labeled.
Solid curve shows the Read-Shockley equation fit ($r_0 = 1.2$~\AA) for
the flat small-angle armchair grain boundaries. Dashed curve shows the 
asymptotic linear dependence of $\gamma$ for the buckled small-angle 
armchair grain boundaries ($E_f =7.5$~eV).
}
\end{figure}

In order to determine the energetically preferred structures and to 
understand the basic thermodynamic properties of grain boundaries in 
graphene, we perform first-principles calculations on models 
containing a pair of complementary dislocations in periodic 2D 
supercell as described in Section~\ref{methods}.
The results are presented as a diagram of grain boundary 
energies per unit length $\gamma$ as a function of $\theta$ 
(Fig.~\ref{fig3}). We first discuss the case of perfectly flat grain 
boundaries (filled symbols) which corresponds to the limit of strong 
binding to a flat substrate and analogous to the case of grain 
boundaries in bulk materials. The diagram clearly reveals both 
armchair and zigzag small-angle regimes with grain boundary formation energies 
converging to zero for $\theta \rightarrow 0^\circ$ and 
$\theta \rightarrow 60^\circ$. A detailed study of the armchair 
small-angle region ($\theta < 10^\circ$) shows that the grain boundary
energies are well described by the Read-Shockley equation\cite{Read50} 
\begin{equation}\label{Read-Shockley}
 \gamma(\theta') = \frac{\mu |\vec{b}|}{4\pi(1-\nu)} \theta' (1+ \ln\frac{|\vec{b}|}{2\pi r_0} -\ln \theta'),
\end{equation}
where $\mu$ is the shear modulus, $\nu$ the Poisson's ratio. The core 
radius $r_0$ encompasses the energy of the dislocation core. In 
Eq.~(\ref{Read-Shockley}), $\theta'=\theta$ or $\theta'=60^\circ - \theta$
for armchair and zigzag small-angle grain boundaries, respectively.
Using the values of elastic constants which correspond to our first 
principles model of graphene (see Appendix~\ref{AppB}) a 
least-squares fit to the Read-Shockley equation (solid red curve in 
Fig.~\ref{fig3}) yields $r_0 = 1.2$~\AA. This value is in good 
agreement with the recently reported $r_0 = 0.96$~\AA\ fitted to 
local density approximation calculations.\cite{Ertekin09} As we 
outlined above, there are several grain boundary structures possible 
for $\theta > 21.8^\circ$. In order to determine the lowest energy 
structure, we compare the energies of grain boundaries constructed from
(1,1) dislocations and (1,0)+(0,1) dislocation pairs. In addition, for
$\theta > 42.1^\circ$ the grain boundary can be constructed either from
equally spaced (1,0) and (0,1) dislocations (disperse case) or from 
closely bound pairs (paired case). Figure~\ref{fig3} shows that the 
disperse (1,0)+(0,1) grain boundaries in flat graphene are the lowest-energy 
structures for $\theta > 42.1^\circ$. More generally, this also 
implies that only the shortest Burgers vector (1,0) dislocation 
is sufficient for constructing the most stable grain boundary structures 
at any given $\theta$.
Remarkably, the two large-angle structures discussed above, LAGB~I and 
LAGB~II, have particularly low formation energies of 0.338~eV/\AA\ 
and 0.284~eV/\AA, respectively. Favorable energetics suggests possible 
abundance of these two structural motifs. Moreover, for all possible values 
of $\theta$ the grain boundary energies are well below the energies of $\sim$1~eV/\AA\ 
predicted for graphene edges.\cite{Koskinen08}

The case of free-standing 2D materials is notably different since 
buckling in the third dimension allows an exchange of in-plane elastic 
energy for bending energy. This leads to efficient screening of the 
in-plane strain field resulting in the finite formation energies of
dislocations.\cite{Seung88} While we find that the large-angle grain 
boundaries in graphene are flat, for $\theta<21.8^\circ$ and $\theta>38.2^\circ$
buckling effectively reduces the grain boundary energies (Fig.~\ref{fig3}, empty symbols). 
In the small-angle regimes, grain boundary energy is
expected to scale linearly with~$\theta'$:
\begin{equation}\label{linear}
 \gamma(\theta') = \frac{E_f \theta'}{|\vec{b}|},
\end{equation}
where $E_f$ is the formation energy of the dislocation. By 
extrapolating the grain boundary energies to $\theta = 0^\circ$, we 
obtain a formation energy of 7.5~eV for the (1,0) dislocation (see 
Appendix~\ref{AppC}). This value is comparable to the formation
energies of typical point defects in graphene, e.g. vacancies (7.6~eV)
and Stone-Wales defects (4.8~eV).\cite{Li05} Out-of-plane buckling 
results in a prolate hillock appearance of dislocations on the flat 
graphene surface (Fig.~\ref{fig2}(f)) in agreement with the experimental 
observations of Coraux {\it et al.}\cite{Coraux08} The height of 
the protrusion around the (1,0) dislocation is $\sim$3~\AA\ and its 
top is shifted with respect to the dislocation core. Interestingly, 
out-of-plane distortion makes the paired case (1,0)+(0,1) grain 
boundaries more stable, thus inverting the sign of effective interaction
between the dislocation dipoles. The situations in which graphene is 
bound to substrate can be viewed as intermediate between the flat and
buckled regimes.

\subsection{Electronic structure of topological defects}\label{electronic}

\begin{figure}
\includegraphics[width=8.5cm]{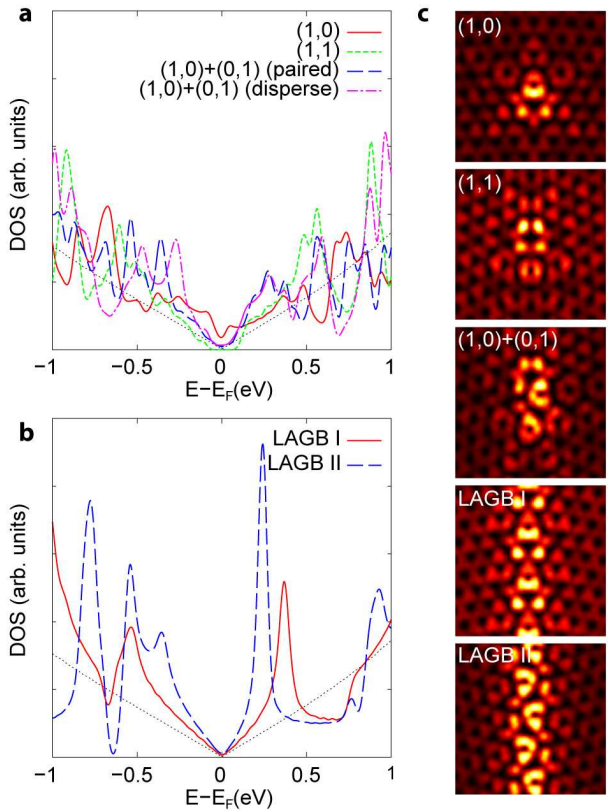}
\caption{\label{fig4} 
(color online).
(a) Calculated density-of-states plots for the small-angle
armchair ($\theta = 7.3^\circ$) and the possible configurations of
small-angle zigzag ($\theta = 49.5^\circ$) 
grain boundaries. The plots correspond to the values averaged over 2~nm 
wide interface regions. The dotted line shows the density-of-states
of the ideal graphene. (b) Calculated density-of-states plots for 
the large-angle grain boundary structures LAGB~I and LAGB~II.
(c) Simulated STM images of the individual dislocations in small-angle
grain boundaries and large-angle structures. The images cover 2~nm$~\times$~2~nm areas.
}
\end{figure}

\begin{figure}
\includegraphics[width=8.5cm]{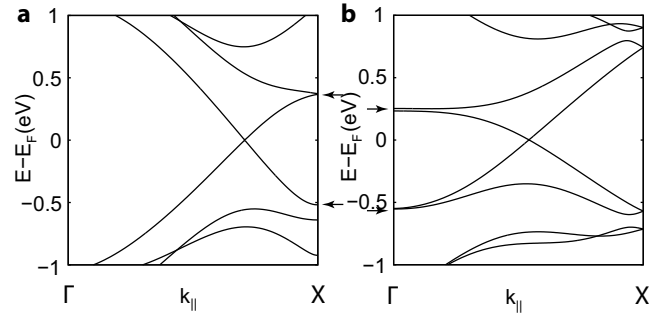}
\caption{\label{figs4}
The calculated band structures for the models of (a) LAGB I and (b) LAGB II 
large-angle grain boundaries along $k_{||}$ at $k_\perp = 0$.
Correspondence between the van Hove singularities in Figure~\ref{fig4}(b) 
and the band extrema in the band structure plots is highlighted with arrows.  
}
\end{figure}

Finally, we address the electronic structure 
of topological defects in graphene. Figures~\ref{fig4}(a) and \ref{fig4}(b) show the 
calculated density-of-states plots for small-angle and stable large-angle
grain boundaries in graphene. All studied defect configurations 
introduce van Hove singularities within 0.5~eV below and above the Dirac point 
($E_F = 0$~eV), in accordance with the scanning tunneling spectroscopy (STS)
observations for the majority of grain boundaries in graphite 
reported in Refs.~\onlinecite{Cervenka09,Cervenka09b}. 
The van Hove singularities are the signatures of one-dimensional states localized 
at the interface as shown by simulated STM images (Fig.~\ref{fig4}(c)).
This relation is further corroborated by considering the electronic band structures 
of large-angle grain boundary models shown in Figure~\ref{figs4}.
However, we do not observe any zero-energy states or defect-induced magnetic moments typical 
of zigzag edges\cite{Son06,Yazyev08b} and single-atom defects\cite{Yazyev07,Yazyev08,Yazyev10} in graphene.
Pronounced changes in the low-energy part of electron spectrum make it possible to
identify the discussed extended defects using STM. In order to facilitate the attribution of 
experimental observations to the proposed structures we provide their 
simulated atomic-scale STM fingerprints (Fig.~\ref{fig4}(c)). The common feature of all images is the crescent- 
or ring-shaped appearance of 5-membered rings. The STM images of grain 
boundaries formed by (1,0)+(0,1) dislocations show the lack of mirror 
symmetry compared to the (1,0) or (1,1) derived structures, as follows 
from the atomic structures of these grain boundaries.

\section{CONCLUSIONS}\label{conclusions}  

We have developed a systematic approach for constructing atomic 
structures of topological defects in graphene. Our first-principles
calculations revealed a number of intriguing features in the energetics
of grain boundaries. In particular, we have found two large-angle 
grain boundary structures with particularly low formation 
energies as well as two distinct small-angle regimes which correspond 
to the grain boundaries oriented close to the armchair and zigzag 
directions, respectively. In free-standing graphene the small-angle 
grain boundaries show pronounced tendency to an out-of-plane
buckling which further reduces their formation energies. We have also
found that all the studied topological defects have strong effects on the electronic 
structure and can be identified using STM. These results show that 
dislocations and grain boundaries are important intrinsic defects in 
graphene which may be used for engineering graphene-based nanomaterials
and functional devices.

\section*{ACKNOWLEDGMENTS}

We are grateful to Y.-W. Son and D. Strubbe for their suggestions. 
This work was supported by National Science Foundation Grant No.~DMR07-05941 
and by the Director, Office of Science, Office of Basic Energy Sciences, 
Division of Materials Sciences and Engineering Division, U.S. Department 
of Energy under Contract No.~DE-AC02-05CH11231.
O.\ V.\ Y. acknowledges financial support of the Swiss National Science 
Foundation (grant No.~PBELP2-123086). 
Computational resources have been provided by NERSC and TeraGrid.

\appendix

\section{Relation between the dislocation dipole $\vec{a}$ and the Burgers vector $\vec{b}$}\label{AppA}

{\bf Theorem.}
{\it If a dislocation is constructed from a pair of $s=\pm \pi/3$ disclinations, then
its Burgers vector $\vec{b}$ and vector $\vec{a}$ connecting the disclinations
are related as}
\begin{equation}
 |\vec{a}|=|\vec{b}|.
\end{equation}

{\it Proof.}
We consider the following construction in which a negative $s=-\pi/3$ disclination 
inserts sector $AHB$ at point $H$ and a complementary positive $s=\pi/3$ disclination
removes sector $A''PB''$ at point $P$ in a continuous two-dimensional sheet. 

\begin{figure}[b]
\includegraphics[width=8.5cm]{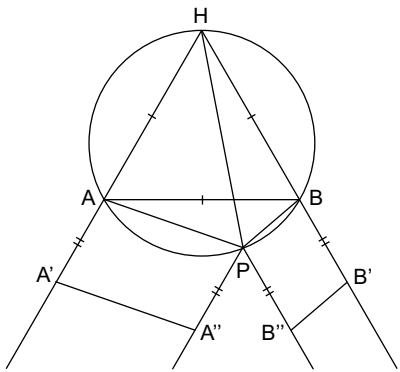}
\caption{\label{figs1}
Sketch of the geometric construction of a dislocation in a continuous sheet. Negative and 
positive $\pi/3$ disclinations are placed at points $H$ and $P$, respectively.}
\end{figure}

The condition of continuity requires $\overline{AH}=\overline{BH}$ ($=\overline{AB}$), 
$\overline{A'H}=\overline{B'H}$ as well as $\overline{A''H}=\overline{B''H}$. In addition, 
we require that, after the described procedure, pairs of segments $A'A''$ and 
$B'B''$ as well as $AP$ and $BP$ form straight lines. That is,
\begin{equation}
 \overline{A'A''}  + \overline{B'B''} = \overline{AP}  + \overline{BP} =|\vec{b}|.
\end{equation}
Hence,
\begin{equation}
 \widehat{APA''} + \widehat{BPB''} = \pi.
\end{equation}

By construction $\widehat{AHB} = \widehat{A''PB''} = \pi/3$, thus $\widehat{APB} = 2\pi/3$ and 
$\widehat{APB} + \widehat{AHB} = \widehat{HAP} + \widehat{HBP} = \pi$.
Due to the latter property, quadrilateral $AHBP$ can be inscribed in a circle 
and, thus, from Ptolemy's theorem it follows that 
\begin{equation}
 \overline{AB} \cdot \overline{HP} = \overline{BH} \cdot \overline{AP} + \overline{AH} \cdot \overline{BP}.
\end{equation}

Finally,
\begin{equation}
 |\vec{a}| = \overline{HP} = \overline{AP} + \overline{BP} = |\vec{b}|.
\end{equation}

\section{Structural and elastic constants of graphene from first principles}\label{AppB}

In order to fit the results of our calculations of small-angle grain 
boundaries to the continuum-model Read-Shockley equation, we use 
elastic constants and the interatomic distance of graphene which
correspond to the present first-principles model of graphene. 
Elastic constants are obtained from the constrained variable cell 
calculations in which one of the rectangular supercell dimensions was 
fixed while the other one varies in order to minimize the total energy.
This allowed us to determine Young's modulus $E$ and Poisson's ratio $\nu$ while 
the shear modulus $\mu$ was calculated using the following relation:\cite{S1}
\begin{equation}
 \mu = \frac{E}{2(1+\nu)}.
\end{equation}
The calculated moduli correspond to 3.35~\AA\ thickness of the graphene layer. 
Our values are in good agreement with other values reported in literature (see Table~\ref{tab1}).

\begin{table}[b]
\caption{\label{tab1} 
The values of interatomic distance $d_{\rm cc}$, Young's 
modulus $E$, Poisson's ratio $\nu$ and the shear modulus $\mu$ of graphene 
obtained from the present first-principles calculations. 
The results are compared to values reported in literature.
}
\begin{ruledtabular}
\begin{tabular}{lcccc}
                     & $d_{\rm cc}$ (\AA) & $E$ (TPa) & $\nu$  & $\mu$ (GPa) \\
\hline
 this work           & 1.433              & 1.052     & 0.206  &    436 \\
 Exp. (graphite) \cite{S2}&                    & 1.02$\pm$0.03 & 0.16$\pm$0.03 & \\
 Exp. (graphene) \cite{S3}&                    & 1.0$\pm$0.1   & & \\  
 Theory          \cite{S4}&                    & 1.050         & 0.186 & \\
 Theory          \cite{S5}&                    & 1.01$\pm$0.03 & 0.21$\pm$0.01 & \\
\end{tabular}
\end{ruledtabular}
\end{table}

\section{Formation energy of the buckled (1,0) dislocation in graphene}\label{AppC}

Figure~\ref{fig3}  shows that the energies of the buckled small-angle 
armchair grain boundaries do not achieve the expected linear dependence 
(Eq. (\ref{linear})) in the range of studied misorientation angles $\theta$. 
The maximum separation $D$ between the neighboring (1,0) dislocations along the
boundary line we could afford in our demanding first-principles calculations
is $\sim$4~nm (corresponds to $\theta =3.5^\circ$). At this $D$, the screening of 
the in-plane elastic field is still insufficient to decouple the neighboring
dislocations along the grain boundary direction. However, we observe that the
energy per dislocation $\gamma D$ shows a clear linear dependence with $1/D$.
By extrapolating the values of grain boundary energies for $\theta < 10^\circ$ 
to the limit of $1/D=0$ (that is, $\theta=0^\circ$), we obtain an estimate of the formation energy 
of an isolated buckled (1,0) dislocation $E_f = 7.5$~eV.

\begin{figure}
\includegraphics[width=8.5cm]{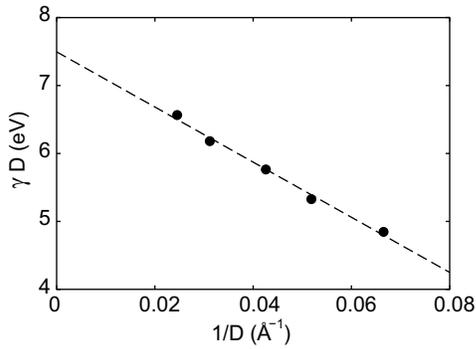}
\caption{\label{figs2}
Grain boundary energies per dislocation $\gamma D$ as a function of inverse 
distance $D$ between the neighboring dislocations. Dashed line shows the least-squares
fit.}
\end{figure}

\end{document}